\documentstyle[preprint,prabib,aps]{revtex}

\begin{document}

\title{Locally Anisotropic Wormholes
and Flux Tubes in 5D Gravity }

\draft
\date{\today}

\author{Sergiu I. Vacaru
\thanks{E-Mails :  sergiu$_{-}$vacaru@yahoo.com,\   
 sergiuvacaru@venus.nipne.ro}}
\address{Institute for Space Sciences,
B. O. Box: MG-23, RO76900, M\v agurele, Bucharest, Romania} 

\author{D. Singleton
\thanks{E-Mail : dougs@csufresno.edu}}
\address{Dept. of Phys., CSU Fresno, 2345 East San Ramon Ave.
M/S 37 Fresno, CA 93740-8031, USA}

\author{Vitalie A. Bo\c tan \thanks{E-Mail : botsan$_{-}$v@yahoo.com}
and Denis A. Do\c tenco \thanks{E-Mail :
dotsencodenis@yahoo.com, dotsenco@mail.com}}
\address{Faculty of Physics, State University of Moldova
60 Mateevici str., Chi\c sin\v au MD2009, Republic of Moldova}

\maketitle

\begin{abstract}
In this article we examine a class of wormhole and flux tube like solutions
to 5D vacuum Einstein equations. These solutions possess generic local
anisotropy, and their local isotropic limit is shown to be conformally
equivalent to the spherically symmetric 5D solutions of Ref. \cite{ds}.
The anisotropic solutions investigated here have two physically
distinct signatures: First, they can give rise to angular--dependent,
anisotropic ``electromagnetic'' interactions.
Second, they can result in a gravitational running of the ``electric''
and ``magnetic'' charges of the solutions.  This gravitational running
of the electromagnetic charges is linear rather than logarithmic,
and could thus serve as an indirect signal for the presence of
higher dimensions. The local anisotropy of these solutions is modeled
using the technique of anholonomic frames with respect to which
the metrics are diagonalized. If holonomic coordinate frames were
used then such metrics would have off--diagonal components.
\end{abstract}

\vspace{1cm} 

\vspace{1cm}

\section{Introduction}

Recently one of the authors \cite{v} studied a class of solutions to
Einstein's equations in three (3D) and four (4D) dimensions which had generic
local anisotropy ({\it e.g.} static black hole and cosmological solutions
with ellipsoidal or toroidal symmetry). These solutions were obtained and
studied using the method of moving anholonomic frames with an associated
nonlinear connection structure on (pseudo) Riemannian spaces. This technique
was formally developed in Refs. \cite{vst} where it was used to investigate
locally anisotropic (super) string theory and supergravity models with
anisotropic structures. In this paper we apply this method of anholonomic
frames to Kaluza--Klein theory in order to construct metrics which describe
locally anisotropic wormhole and flux tube solutions for the 5D vacuum
Einstein equations. 

In 4D and higher dimensional gravity there are well known spherical
symmetric solutions which describe black holes and wormholes \cite{mor}.
These solutions have diagonal metrics. Considering metrics in higher
dimensional gravity which have non--zero off--diagonal components can lead
to interesting physical consequences. Salam, Strathee and Perracci \cite{sal}
showed that including off--diagonal components in higher dimensional metrics
is equivalent to including gauge fields,
and concluded that geometrical gauge fields can act as possible sources of
exotic matter necessary for the construction of a wormhole.

In Refs. \cite{chodos,dzhsin} various locally isotropic solutions with
off--diagonal metric components were obtained for 5D vacuum Einstein
equations. These solutions were similar to spherically symmetric 4D wormhole
and/or flux tube metrics with ``electric'' and/or ``magnetic'' fields running
along the throat of the wormhole. These ``electromagnetic'' fields arose as a
consequence of the off--diagonal elements of the metric. By varying certain
free parameters of the off--diagonal elements of the 5D metric, it was
possible to change the relative strengths of the fields in the wormholes
throat, and to change the longitudinal and transverse size of the wormhole's
throat. Also for certain values of these parameters the 5D solutions could
be related to 4D Reissner--Nordstrom solutions.

In the present work we use the method of anholonomic frames with associated
nonlinear connections to construct locally {\it anisotropic}
wormhole and flux tube solutions that reduce to the solutions of
Ref. \cite{ds} in the isotropic limit. These solutions have
off--diagonal (with respect to coordinate frames) metrics with
generic, anholonomic vacuum polarizations of the 5D vacuum
gravitational fields. These solutions exhibit two, distinct
features depending on how the anisotropies are introduced.
For anisotropies in the extra spatial dimension, $\chi$, the solutions
have a gravitational running or scaling of the
``electromagnetic'' charges. This gravitational scaling is linear rather
than logarithmic, and might therefore provide an indirect
signature for the presence of extra dimensions. For anisotropies
in the axial angle, $\varphi$, we find that an angle--dependent, anisotropic
``electromagnetic'' interaction results for a test particle which
moves in the background field of the solution.

The paper is organized as follows : In Sec. 2 we review the 
results for locally isotropic wormhole and flux tube solutions. In Sec. 3 we
present the necessary geometric background on anholonomic frames and
associated nonlinear connections in 5D (pseudo) Riemannian geometry. In Sec.
4 we construct and analyze a new class of 5D metrics describing anisotropic
wormhole and flux tube solutions of the vacuum Einstein equations. 
In Sec. 5 we give an example of a wormhole like solution
where the extra spatial dimension is handled by an exponential
warp factor \cite{rs} rather than being compactified as in
standard Kaluza-Klein theory. In Sec. 6 we summarize
and discuss the main conclusions of the paper.

\section{ 5D Locally Isotropic Wormhole Like Solutions}

Here we give a brief review of the wormhole and flux tube solutions
(DS-solutions) of Refs. \cite{ds,dzhsin}.
The ansatz for the spherically symmetric 5D metric is taken as
\begin{eqnarray}
ds^2_{(DS)} &=& e^{2\nu (r)}dt^{2} -
 dr^{2} - a(r)(d\theta ^{2} + \sin ^{2}\theta  d\varphi ^2)
\nonumber \\
& -& r_0^2e^{2\psi (r) - 2\nu (r)}
\left [d x^5 +  \omega (r)dt + n\cos \theta d\varphi \right ]^2,
\label{ansatz1}
\end{eqnarray}
$x^5 $ is the 5$^{th}$ extra, spatial coordinate;
$r,\theta ,\varphi $ are $3D$ spherical coordinates;
$n$ is an integer; $r\in \{-R_0,+R_0\}$ ($R_0\leq $ 
$\infty $) and $r_0$ is a constant. All functions $\nu (r),\psi (r)$ and
$a(r)$ are taken to be even functions of $r$ satisfying $\nu ^{\prime
}(0)=\psi ^{\prime }(0)=a^{\prime }(0)=0$. The coefficient $\omega (r)$ is
treated as the $t$--component of the electromagnetic potential and $n\cos
\theta $ is the $\varphi $-component. These electromagnetic potentials lead
to this metric having radial Kaluza-Klein ``electrical'' and ``magnetic''
fields. The metric (\ref{ansatz1}) is assumed to solve the 5D Einstein vacuum
equations
\begin{equation}
\label{5einstein}R_{\alpha \beta }-\frac 12g_{\alpha \beta }R=0,
\end{equation}
Greek indices $\alpha ,\beta ,...$ run over $0,1,2,3,4$. The 5D Kaluza-Klein
``electric'' field is
\begin{equation}
\label{emf}E_{KK}=r_0\omega ^{\prime }e^{3\psi -4\nu }=
\frac{q_0}{a(r)}
\end{equation}
with the ``electric'' charge
\begin{equation}
\label{emfec}q_0=r_0\omega ^{\prime }(0)
\end{equation}
which can be parametrized as
\begin{equation}
\label{emfeca}q_0=2\sqrt{a(0)}\sin \alpha _0,
\end{equation}
The corresponding dual, ``magnetic'' field $H_{KK} = Q_0/a(r)$ with
``magnetic'' charge $Q_0=nr_0$ is parametrized as
$$
Q_0=2\sqrt{a(0)}\cos \alpha _0,
$$
The following relation
\begin{equation}
\label{emfm}
\frac{(q_0^2+Q_0^2)}{4a(0)}{=1}
\end{equation}
relates the ``electric'' and ``magnetic'' charges. Later when we
consider anisotropic versions of the metric in Eq. (\ref{ansatz1})
the charges will have a dependence on $\varphi$ or $\chi$.
Eq. (\ref{emfm}) will still hold for these coordinate
dependent charges, and thus enforces a definite relationship
between the ``electric'' and ``magnetic'' charges.
The solution in \cite{ds} satisfied the boundary conditions
$a(0)=1,\psi (0)=\nu (0)=0.$ As the free parameters of the metric
are varied there are five classes of solutions with the properties:

\begin{enumerate}
\item  $Q=0$ or $H_{KK}=0$, a wormhole--like ``electric'' object;

\item  $q=0$ or $E_{KK}=0$, a finite ``magnetic'' flux tube;

\item  $H_{KK}=E_{KK}$, an infinite ``electromagnetic'' flux tube;

\item  $H_{KK}<E_{KK}$, a wormhole--like ``electromagnetic'' object;

\item  $H_{KK}>E_{KK}$, a finite, ``magnetic--electric'' flux tube.
\end{enumerate}

Here we will generalize these solutions (\ref{ansatz1}) to locally
anisotropic configurations using anholonomic frames.

In order to simplify the procedure of construction of new classes
of solutions with generic local anisotropy it convenient to introduce
a new 5$^{th}$ coordinate $\chi$ by a transform  $x^5 \to \chi$ so
that the coordinates are related via
$$
dx^5 +  n\cos \theta d\varphi =
d\chi +  n\cos \theta d\theta
$$
this can be accomplished by taking
$\chi = x^5 - \int [\mu (\theta, \varphi)]^{-1}
d\xi (\theta, \varphi)$
with 
$$
\frac {\partial \xi}{\partial \varphi}
= - \mu n\cos \theta, \qquad 
\frac {\partial \xi}{\partial \theta}
= \mu n\cos \theta
$$
and
$$
\mu (\theta, \varphi) =
\exp (\theta  - \varphi)|\cos \theta | ^{-1}.
$$
so that the mixed partials of $\xi$ are consistent.
With respect to the new extra dimensional coordinate
$\xi$ the $A_{\varphi}$ component of the 
electromagnetic potential is exchanged for 
a component $A_{\theta}$ and the metric
interval (\ref{ansatz1}) is rewritten
as
\begin{eqnarray}
ds^2_{(DS,new)} &=& e^{2\nu (r)}dt^{2} -
dr^{2} - a(r)(d\theta ^{2} + \sin ^{2}\theta  d\varphi ^2)
\nonumber \\
& -& r_0^2e^{2\psi (r) - 2\nu (r)}
\left [d \chi +  \omega (r)dt + n\cos \theta d\theta \right ]^2,
\label{ansatz1n}
\end{eqnarray}
This allows us to treat the $(t,r,\theta)$
coordinates as holonomic and the
$(\varphi, \chi)$ coordinates  as anholonomic.

\section{Anholonomic Frames and Local Anisotropy}

Let us consider a 5D pseudo--Riemannian
spacetime of signature $(+,-,-,-,-)$
and denote the local coordinates $u^\alpha =(x^i,y^a),$ -- or more
compactly $u=(x,y)$ -- where the Greek indices are conventionally split into
two subsets $x^i$ and $y^a$ labeled respectively by Latin indices of type
$i,j,k,...=1,2,...n$ and $a,b,...=1,2,...,m,$ with $n+m=5.$ The local
coordinate bases, $\partial _\alpha =(\partial _i,\partial _a),$ and their
duals, $d^\alpha =\left( d^i,d^a\right) ,$ are defined respectively as
\begin{equation}
\label{pder}\partial _\alpha \equiv \frac \partial {du^\alpha }=
(\partial_i=\frac \partial {dx^i},\partial _a=\frac \partial {dy^a})
\end{equation}
and
\begin{equation}
\label{pdif}d^\alpha \equiv du^\alpha =(d^i=dx^i,d^a=dy^a).
\end{equation}
In Kaluza--Klein theories (see Ref. \cite{over} for a review) one often
parameterizes the metric as
\begin{equation}
\label{ansatz2}g_{\alpha \beta }=\left[
\begin{array}{cc}
g_{ij}(u)+N_i^a(u)N_j^b(u)h_{ab}(u) & N_j^e(u)h_{ae}(u) \\
N_i^e(u)h_{be}(u) & h_{ab}(u)
\end{array}
\right]
\end{equation}
with the coefficients, $N_i ^a (u)$ {\it etc.}, given with respect to a
local coordinate basis. A 5D metric $g^{(5)}$ with coefficients (\ref
{ansatz2}) splits into a block $\left( n\times n\right) +\left( m\times
m\right) $ form
\begin{equation}
\label{dmetr1}\delta s^2=g_{ij}\left( x,y\right) dx^idx^j+h_{ab}\left(
x,y\right) \delta y^a\delta y^b
\end{equation}
Instead of using the coordinate bases (\ref{pder}) and (\ref{pdif}) one can
introduce locally anisotropic frames (bases)
\begin{equation}
\label{dder}\delta _\alpha \equiv \frac \delta {du^\alpha }=(\delta
_i=\partial _i-N_i^b(u)\ \partial _b,\partial _a=\frac \partial {dy^a})
\end{equation}
and
\begin{equation}
\label{ddif}\delta ^\alpha \equiv \delta u^\alpha =(\delta ^i=dx^i,\delta
^a=dy^a+N_k^a(u)\ dx^k).
\end{equation}

The main `trick' of the anholonomic frames method for integrating the
Einstein equations \cite{v,vst} is to find $N^a_j$'s such that the block
matrices $g_{ij}$ and $h_{ab}$ are diagonalized. This greatly simplifies
computations. With respect to these anholonomic frames the partial
derivatives are N--elongated (locally anisotropic).

This topic of anholonomic frames is related to the geometry of moving
frames and nonlinear connection structures on manifolds and vector bundles.
The idea is to take the coefficients $N_i^a(x,y)$ as defining a nonlinear
connection (an N--connection) structure with the curvature
(N--curvature)
$$
\Omega _{ij}^a\doteq \delta _jN_i^a-\delta _iN_j^a
$$
This induces a global decomposition of the 5D pseudo--Riemannian spacetime
into holonomic (horizontal, h) and anholonomic (vertical, v) directions. In
a preliminary form the concept of N--connections was applied by E. Cartan in
his approach to Finsler geometry \cite{cartan} and a rigorous definition was
given by Barthel \cite{barthel} (Ref. \cite{ma} gives a modern approach to
the geometry of N--connections, and to generalized Lagrange and Finsler
geometry). As a particular case one obtains the linear connections if 
$N_i^a(x,y)=\Gamma _{bi}^a\left( x\right) y^a.$

A more surprising result is that N--connection structures can be naturally
defined on (pseudo) Riemannian spacetimes \cite{v} by associating them with
some anholonomic frame fields (vielbeins) of type (\ref{dder}) satisfying
the relations
$$
\delta _\alpha \delta _\beta -\delta _\beta \delta _\alpha =w_{\ \alpha
\beta }^\gamma \delta _\gamma
$$
where the anholonomy coefficients $w_{\ \alpha \beta }^\gamma $ are computed
as follows
\begin{eqnarray}
w_{\ ij}^k &=& 0 \; , \; w_{\ aj}^k=0 \; , \;
w_{\ ia}^k=0 \; , \; w_{\ ab}^k=0 \; , \; w_{\ ab}^c=0,
\nonumber \\
w_{\ ij}^a &=&-\Omega _{ij}^a \; , \; \ w_{\ bj}^a=-\partial _bN_j^a \; ,\;
 w_{\ ia}^b= \partial _a N^b_i. \label{anhol}
\end{eqnarray}
One says that the N--connection coefficients model a locally anisotropic
structure on spacetime ( an locally anisotropic 
spacetime) when the partial derivative
operators and coordinate differentials, (\ref{pder}) and (\ref{pdif}), are
respectively changed into N--elongated operators (\ref{dder}) and
(\ref{ddif}).

A linear connection $D_{\delta _\gamma }\delta _\beta =\Gamma _{\ \beta
\gamma }^\alpha \left( x,y\right) \delta _\alpha $ is compatible with the
metric $g_{\alpha \beta }$ and N--connection structure on a 5D
pseudo--Riemannian spacetimes, if $D_\alpha g_{\beta \gamma }=0.$ The linear
connection is parametrized by irreducible h--v--components,
$$
\Gamma _{\ \beta \gamma }^\alpha =\left( L_{\ jk}^i,L_{\ bk}^a,C_{\
jc}^i,C_{\ bc}^a\right) ,
$$
where
\begin{eqnarray}
L_{\ jk}^i &=&\frac 12g^{in}\left( \delta _kg_{nj}+\delta _jg_{nk}-\delta
_ng_{jk}\right) , \nonumber \\
L_{\ bk}^a &=&\partial _bN_k^a+\frac 12h^{ac}\left( \delta
_kh_{bc}-h_{dc}\partial _bN_k^d-h_{db}\partial _cN_k^d\right), 
\label{dcon}  \\
C_{\ jc}^i &=&\frac 12g^{ik}\partial _cg_{jk},\  
C_{\ bc}^a = \frac 12h^{ad}\left( \partial _ch_{db}+
\partial _bh_{dc}-\partial _dh_{bc}\right) \nonumber
\end{eqnarray}
This defines a canonical linear connection (as distinguished from an
N--connection) which is similar to the metric connection introduced by
Christoffel symbols in the case of holonomic bases.

The anholonomic coefficients $w_{\ \alpha \beta }^\gamma $ and N--elongated
derivatives give nontrivial coefficients for the torsion tensor, $T(\delta
_\gamma ,\delta _\beta )=T_{\ \beta \gamma }^\alpha \delta _\alpha ,$ where
\begin{equation}
\label{torsion}T_{\ \beta \gamma }^\alpha =\Gamma _{\ \beta \gamma }^\alpha
-\Gamma _{\ \gamma \beta }^\alpha +w_{\ \beta \gamma }^\alpha ,
\end{equation}
and for the curvature tensor, $R(\delta _\tau ,\delta _\gamma )\delta _\beta
=R_{\beta \ \gamma \tau }^{\ \alpha }\delta _\alpha ,$ where
\begin{equation}
R_{\beta \ \gamma \tau }^{\ \alpha } =
\delta _\tau \Gamma _{\ \beta \gamma}^\alpha
-\delta _\gamma \Gamma _{\ \beta \tau }^\alpha  \label{curvature}
+ \Gamma _{\ \beta \gamma }^\varphi \Gamma _{\ \varphi \tau }^\alpha -
\Gamma _{\ \beta \tau }^\varphi \Gamma _{\ \varphi \gamma }^\alpha +
\Gamma _{\ \beta \varphi }^\alpha w_{\ \gamma \tau }^\varphi .
\end{equation}
We emphasize that the torsion tensor on (pseudo) Riemannian spacetimes is
induced by anholonomic frames, whereas its components vanish with respect to
holonomic frames. All tensors are distinguished (d) by the N--connection
structure into irreducible h--v--components, and are called d--tensors. For
instance, the torsion, d--tensor has the following irreducible,
nonvanishing, h--v--components,
$$
T_{\ \beta \gamma }^\alpha =\{T_{\ jk}^i,C_{\ ja}^i,S_{\ bc}^a,T_{\
ij}^a,T_{\ bi}^a\},
$$
where
\begin{eqnarray}
T_{.jk}^i & = & T_{jk}^i=L_{jk}^i-L_{kj}^i,\quad
T_{ja}^i=C_{.ja}^i , \quad T_{aj}^i=-C_{ja}^i, T_{.ja}^i  =  0,\
 \nonumber \\
 T_{.bc}^a &=& S_{.bc}^a=C_{bc}^a-C_{cb}^a,
T_{.ij}^a  = -\Omega _{ij}^a,\quad T_{.bi}^a= \partial _b  N_i^a
-L_{.bi}^a,\quad T_{.ib}^a=-T_{.bi}^a \nonumber
\end{eqnarray}
(the d--torsion is computed by substituting the h--v--components of the
canonical d--connection (\ref{dcon}) and anholonomic coefficients (\ref
{anhol}) into the formula for the torsion coefficients (\ref{torsion})). 
The curvature d-tensor has the following irreducible, non-vanishing,
h--v--components
$$
R_{\beta \ \gamma \tau }^{\ \alpha
}=\{R_{h.jk}^{.i},R_{b.jk}^{.a},P_{j.ka}^{.i},P_{b.ka}^{.c},
S_{j.bc}^{.i},S_{b.cd}^{.a}\},
$$
where
\begin{eqnarray}
R_{h.jk}^{.i} &=&  \delta _kL_{.hj}^i-\delta_jL_{.hk}^i 
+  L_{.hj}^mL_{mk}^i-L_{.hk}^mL_{mj}^i-C_{.ha}^i\Omega _{.jk}^a,
\nonumber \\
R_{b.jk}^{.a} &=&  \delta _kL_{.bj}^a-\delta_jL_{.bk}^a 
+  L_{.bj}^cL_{.ck}^a-L_{.bk}^cL_{.cj}^a-C_{.bc}^a\Omega _{.jk}^c,
\nonumber \\
P_{j.ka}^{.i} &=&  \partial _aL_{.jk}^i +C_{.jb}^iT_{.ka}^b 
-  ( \delta _kC_{.ja}^i+L_{.lk}^iC_{.ja}^l -
L_{.jk}^lC_{.la}^i-L_{.ak}^cC_{.jc}^i ), \nonumber \\
P_{b.ka}^{.c} &=&  \partial _aL_{.bk}^c +C_{.bd}^cT_{.ka}^d 
- ( \delta _kC_{.ba}^c+L_{.dk}^{c\ }C_{.ba}^d
- L_{.bk}^dC_{.da}^c-L_{.ak}^dC_{.bd}^c ), \nonumber \\
S_{j.bc}^{.i} &=&  \partial _cC_{.jb}^i-\partial _bC_{.jc}^i
 +  C_{.jb}^hC_{.hc}^i-C_{.jc}^hC_{hb}^i, \nonumber \\
S_{b.cd}^{.a} &=& \partial _dC_{.bc}^a-\partial
_cC_{.bd}^a+C_{.bc}^eC_{.ed}^a-C_{.bd}^eC_{.ec}^a \nonumber
\end{eqnarray}
(the d--curvature components are computed in a similar fashion by using the
formula for curvature coefficients (\ref{curvature})).
The Ricci tensor $R_{\beta \gamma }=R_{\beta ~\gamma \alpha }^{~\alpha } $
has the d--components
\begin{equation} \label{dricci} 
R_{ij}  =  R_{i.jk}^{.k},\quad
 R_{ia}=-^2P_{ia}=-P_{i.ka}^{.k}, \qquad
R_{ai} =  ^1P_{ai}=P_{a.ib}^{.b},
\quad R_{ab}=S_{a.bc}^{.c}. 
\end{equation}
In general, since $^1P_{ai}\neq ~^2P_{ia}$, the Ricci d-tensor is
non-symmetric (this could be with respect to anholonomic frames of
reference). The scalar curvature of the metric d--connection,
$\overleftarrow{R}=g^{\beta \gamma }R_{\beta \gamma },$ is computed as
\begin{equation}
\label{dscalar}{\overleftarrow{R}}=G^{\alpha \beta }R_{\alpha \beta }=
\widehat{R}+S,
\end{equation}
where $\widehat{R}=g^{ij}R_{ij}$ and $S=h^{ab}S_{ab}.$

By substituting (\ref{dricci}) and (\ref{dscalar}) into the 5D Einstein
equations (\ref{5einstein}) we obtain a system of vacuum, gravitational
field equations with mixed holonomic--anholonomic degrees of
freedom with an N--connection structure,
\begin{eqnarray}
R_{ij}-\frac 12\left( \widehat{R}+S\right) g_{ij} & = &0, \label{einsteq2} \\
S_{ab}-\frac 12\left( \widehat{R}+S\right) h_{ab} & = & 0, \nonumber \\
^1P_{ai}  =  0,\  ^2P_{ia} & = & 0. \nonumber
\end{eqnarray}
The definition of matter sources with respect to locally 
 anisotropic frames is dealt with in
Refs. \cite{v}. In this paper we deal only with vacuum 5D, locally,
anisotropic gravitational equations.

\section{5D Anisotropic Wormhole Like Solutions}

The 5D metric (\ref{ansatz1n}) lends itself very naturally to the funfbein
(5D frame) formalism. We are interested in constructing and investigating
generalizations of this metric to anisotropic configurations with two
anholonomic coordinates : $\varphi $ and $\chi .$ An ansatz of type (\ref
{dmetr1}), written in terms of an locally anisotropic 
bases of type (\ref{ddif}), is taken
as
\begin{equation}
\label{ansatz3}
\delta s^2 = dt^2-b\left( \zeta \right) d\zeta ^2-
c(\zeta)d\theta ^2
-h_4\left(\zeta,\theta, y^1\right) (\delta y^1)^2 -
 h_5\left( \zeta,\theta, y^1\right) (\delta y^2)^2
 \nonumber
\end{equation}
for $x^1= t,  x^2=\zeta, x^3=\theta , y^1 = v =\varphi 
 (\mbox { or } v = \chi ), y^2=p=\chi 
 (\mbox{ or } p = \varphi ). $
For nontrivial values of $\nu (r)$ the coordinate
$r$ is related to $\zeta$ via $\zeta = \int exp[-\nu(r)] dr$. 
If  $\nu (r)$ is fixed ({\it e.g.} $\nu (r)=0$) then
we can set $\zeta = r$. The N--elongated differentials are
\begin{eqnarray}
\delta  y^1 & = & dv+w_2\left( \zeta, \theta, v\right) d\zeta +
 w_3\left(\zeta ,\theta, v\right) d\theta , \label{ndiff2} \\
\delta y^2 & = & dp+n_2\left( \zeta ,\theta, v\right) d\zeta +
n_3\left(\zeta ,\theta, v\right) d\theta , \nonumber
\end{eqnarray}
{\it i.e.} the coefficients of the N--connection are parametrized
 as $N_1^1
= 0,N_{2,3}^1=w_{2,3}\left( \zeta ,\theta ,v\right), N_1^2 =
0,N_{2,3}^2=n_{2,3}\left( \zeta ,\theta ,v\right).$ We will consider
anisotropic dependencies on the coordinate $y^1=v$
 (a similar construction
holds if we consider dependencies on $y^2=p$).

The nontrivial components of the 5D vacuum Einstein equations
(\ref{einsteq2}) for the metric (\ref{ansatz3}) are computed as
\begin{eqnarray}
\label{einsteq3a}
c^{\prime \prime }-\frac 1{2c} {c^{\prime }}^2
-\frac 1{2b} c^{\prime }b^{\prime } &= &0,\\
{\ddot h}_5 - \frac 1{2h_5} {\dot h_5}^2-\frac 1{2h_4}{ \dot h}_5
{\dot h}_4  & = & 0, \label{einsteq3b} \\
 \beta w_{2,3}+\alpha _{2,3} & = & 0, \label{einsteq3c} \\
\ddot n_{2,3}+\gamma \dot n_{2,3} & = &0, \label{einsteq3d}
\end{eqnarray}
where
\begin{eqnarray} \label{coef2} 
\alpha _2 &=&
\frac{{\dot h}_5}2\left( \frac{h^{\prime }_4}h_4+
\frac{h_5^{\prime }}h_5\right) -{\dot h}_5^{\prime },\qquad  
\alpha _3 = \frac{{\dot h}_5}2\left( \frac{h^{*}_4}h_5+
\frac{h^{*}_5}{h_5}\right) -{\dot h}_5^{*}, \nonumber \\
 \beta &=&
{\ddot h}_5-\frac{{\dot h}_5^2}{2h_5}-\frac{{\dot h}_4} 
{{\dot h}_5}{2h_4},\qquad   
\gamma = \frac 32\frac{{\dot h}_5}{h_5}
-\frac{{\dot h}_4}{h_4}. \nonumber
\end{eqnarray}
For simplicity, the partial derivatives are denoted as
${\dot h}_4=\partial
h_4/\partial v,h_5^{\prime }=\partial h_5/\partial \zeta $ and 
$h_5^{*}=\partial h_5/\partial \theta .$

\subsection{Anisotropic solutions of 5D vacuum Einstein equations}

It is possible to find the general solutions of the Eqs. (\ref
{einsteq3a}), (\ref{einsteq3b}) and ( \ref{einsteq3d}).
Eq. (\ref{einsteq3a}) relates two functions $c(\zeta )$ and $b(\zeta ).$
For a prescribed value of $b(\zeta )$ the general solution is
\begin{equation}
\label{ccoef1}
c(\zeta )=\left[ c_0+c_1\int \sqrt{|b(\zeta )|}d\zeta \right] ^2
\end{equation}
where $c_0$ and $c_1$ are integration constants, which could depend
parametrically on the $\theta$ variable. Alternatively for a
prescribed $c=c(\zeta )$ the general solution is
\begin{equation} \label{bcoef1}
b(\zeta )=b_0[c^{\prime }(\zeta )]^2/c(\zeta )
\end{equation}
with $b_0=const.$ In a similar fashion we can construct the
general solution of (\ref{einsteq3b}),
\begin{equation} \label{fcoef}
h_5\left( \zeta ,\theta ,v\right) =
\left[ h_{5[1]}(\zeta ,\theta )+h_{5[0]}(\zeta
,\theta )\int \sqrt{|h_4(\zeta ,\theta ,v)|}dv\right] ^2,
\end{equation}
where $h_{5[0]}(\zeta ,\theta )$ and $h_{5[1]}(\zeta ,\theta )$ 
are functions of
holonomic variables, or
\begin{equation}
\label{hcoef1}
h_4\left( \zeta ,\theta ,v\right) =h_{4[0]}\left( \zeta ,\theta
\right) [{\dot h}_5(\zeta ,\theta ,v)]^2/h_5(\zeta ,\theta ,v),
\end{equation}
where $h_{5[0]}\left( \zeta ,\theta \right) $ is defined 
from some compatibility
conditions with the local isotropic limit.

Having defined the functions $h_5\left( \zeta ,\theta ,v\right) ,$ and
$h_4\left( \zeta ,\theta ,v\right) $ we can compute
$\alpha_{2, 3}$ and  $\beta$, 
and express the solutions of equations (\ref{einsteq3c}) as
\begin{equation} \label{wcoef}
w_{2,3}=-\alpha _{2,3}/\beta .
\end{equation}

In a similar fashion, after two integrations of the anisotropic
coordinate $v$ we can find the general solution to equations
(\ref{einsteq3d})
\begin{equation} \label{ncoef}
n_{2,3}\left( \zeta ,\theta ,v\right) =
n_{2,3[0]}\left( \zeta ,\theta \right) +
n_{2,3[1]}\left( \zeta ,\theta \right) \int 
 \left( h_4(\zeta ,\theta ,v) / {h_5}^{3/2}(\zeta ,\theta,v)
 \right) dv,
\end{equation}
where $n_{2,3[0]}\left( \zeta ,\theta \right) $ and
 $n_{2,3[1]}\left( \zeta ,\theta \right) $
are defined from boundary conditions.

Equations (\ref{einsteq3c}) are second order linear differential equations
of the anisotropic coordinate $v$. The functions $h_{4, 5}$
can be thought of as parameters; the explicit form of
the solutions of (\ref{einsteq3c}) depends on the values of
these parameters. Compatibility 
with the locally isotropic limit ({\it i.e.} when
$w_{2,3}\rightarrow 0,$) is possible if the conditions $\alpha _{2,3}=0$
are satisfied for isotropic configurations. Methods for constructing
solutions to such equations, when boundary conditions at some $v=v_0$ are
specified, can be found in \cite{kamke}. We will assume that we can always
define the functions $w_{2,3}\left( \zeta ,\theta ,v\right) $ for given
$h_5\left( \zeta ,\theta ,v\right) $ and 
$h_4\left( \zeta ,\theta ,v\right)$;
in this case equations (\ref{einsteq3c}) can be treated as linear
algebraic equations for $w_{2,3}$ with the coefficients 
$\alpha_{2,3}$ and $\beta$ determined using the solutions of
(\ref{einsteq3b}).

\subsection{Locally anisotropic generalizations of DS--metrics}

We now consider two particular types of solutions, which in the locally
isotropic limit can be connected with the DS--metric (\ref{ansatz1n}).
These anisotropic solutions of the 5D vacuum Einstein equations will satisfy
\begin{equation}
b(\zeta )=1,\quad c(\zeta )=a(\zeta ),
\quad h_4(\zeta ,\theta )=
a(\zeta )\sin ^2\theta , \label{aux10}
\end{equation}
as do the locally isotropic solutions of (\ref{ansatz1}),
but now $r_0$ is not a constant, but is a function
like $\widehat{r}_0^2(v).$ The function $h_5$
is parametrized as
\begin{equation}
h_5=\exp [2\psi (\zeta )]\widehat{r}_0^2(v) \label{h5}
\end{equation}
where
\begin{equation}
\label{linr}
\widehat{r}_0^2(v)=r_{0(0)}^2 (1+\varepsilon v) ^2
\end{equation}
The ``renormalization'' constant, $\varepsilon $, is obtained
either from experiment or from some quantum gravity model.
In the locally isotropic limit, ($\varepsilon v\rightarrow 0$)
we recover the DS--metric.

In order to analyze the locally isotropic limit
($\varepsilon v\rightarrow 0$) where the frame functions
transforms like
$$
a(\zeta )\rightarrow a(r),\quad \nu (\zeta )\rightarrow 
\nu (r),\quad \psi
(\zeta )\rightarrow \psi(r)
$$
we impose the limiting condition 
$[{\dot h}_5 (\zeta,\theta ,v)]^2/h_5(\zeta,\theta ,v)
\to \exp[2\psi(r)]$
in (\ref{hcoef1}) which is satisfied for $\varepsilon v \to 0$ if 
$\varepsilon = 1/2r^2_{0(0)}$ and choose the function 
$h_{4[0]} (\zeta,\theta)$ from (\ref{hcoef1}) so that
$$
h_{4[0]} (\zeta,\theta) \exp[2\psi(\zeta)]\to a(r) \sin ^2{\theta} .
$$
It should be emphasized that  (\ref{ccoef1})
gives a dependence like
$$
c(r)=a(r)\exp [\nu (r)]=\left( c_0+c_1r\right) ^2
$$
which holds for a fixed conformal factor $\exp [\nu (r)]$.
Thus in the locally isotropic limit metric (\ref{ansatz3}) becomes 
the DS--metric (\ref{ansatz1n})
multiplied by a conformal factor, $\exp [-2\nu (r)]ds_{(DS)}^2.$

The 5D gravitational vacuum polarization renormalizes the
charge defined in Eq. (\ref{emfec}) as
\begin{equation}
\label{linr2}
q(v)=\widehat{r}_0(v)\omega ^{\prime }(0) =
r_{0(0)} (1+\varepsilon v) \omega ^{\prime }(0).
\end{equation}
The angular parametrization (\ref{emfeca}) also becomes locally
anisotropic,
\begin{equation} \label{charge}
q(v)=2\sqrt{a(0)}\sin \alpha (v),
\end{equation}
and the formula for the ``electric'' field (\ref{emf}) transforms into
\begin{equation} \label{ef}
E_{KK}=\frac{q(v)}{a(r)}.
\end{equation}
For the case when $v=\chi$ Eq. (\ref{linr2}) implies that
the ``electric'' charge of this solution exhibits a linear scaling
with respect to the extra spatial coordinate. This can be
contrasted with the standard quantum field theory calculation where
the electric coupling grows logarithmically as the distance scale
decreases. Also in the present case the scaling of the ``electric'' charge
arises classically as a result of the anisotropy from the extra
spatial coordinate. In contrast the usual logarithmic scaling
of electric charge is a {\it quantum}
field theory result. In the standard Kaluza-Klein
scenario the extra spatial coordinate is taken
to be compactified on the order
of the Planck length ({\it i.e.} $v =\chi \simeq L_{Planck}$).
Unless $\varepsilon$ is unnaturally large
this implies $\varepsilon \chi <<1$ making this linear running of the
``electric'' charge unobservable. However, if the extra dimension(s)
have a larger size \cite{ahamed} then such a linear, gravitational running
of the charges, with respect to the size of the extra
dimension(s), could be observable, and could provide an
indirect indication of the presence of higher dimensions. 
The renormalization of the magnetic charge, $Q_0\rightarrow Q(v)$, can be
obtained using the renormalized ``electric'' charge in relationship (\ref
{emfm}) and solving for $Q(v).$ The form of (\ref{emfm}) implies that the
running of the ``magnetic'' charge $Q(v)$ will be the opposite that of the
``electric'' charge, $q(v)$. For example, if $q(v)$ increases with $v$ then 
$Q(v)$ will decrease. For small values of the ``renormalization''
constant, $\varepsilon$, the function $a(r)$ will take a form similar
to that of the DS--solution.

In the case were the anisotropy arises from the $\varphi$ coordinate
one can see that the ``electric'' charge in Eq. (\ref{linr2}) has
a dependence on $\varphi$, which leads to the Kaluza-Klein electric
field having an anisotropic $\varphi$ dependence. A test charge placed
in the background of such a solution would experience an anisotropic,
$\varphi$--dependent interaction. Thus the anisotropies of this solution
result in anisotropic interactions for the matter fields in 4D.

The usefulness of the anholonomic frames method in dealing
with anisotropic configurations can be seen by writing out
the ansatz metric (\ref{dmetr1}), for the two classes of
solutions ({\it i.e.} $v=\varphi$ or $v=\chi$) discussed above,
in a coordinate frame (\ref{pder}). 
\begin{equation} \label{ansatz4}
\left[
\begin{array}{ccccc}
1 & 0 & 0 & 0 & 0 \\
0 & -1-w_2^{\ 2}h_4 -n_2^{\ 2}h_5 & -w_2 w_3 h_4  
-n_2 n_3 h_5 & -w_2h_4 & -n_2h_5 \\
0 & -w_3 w_2 h_4 - n_2 n_3 h_5 & -a-w_3^{\ 2}h_4 
-n_3^{\ 2}h_5 & -w_3h_4 & -n_3h_5 \\
0 & -w_2h_4 & -w_3h_4 & -h_4 & 0 \\
0 & -n_2h_5 & -n_3h_5 & 0 & -h_5
\end{array}
\right] . 
\end{equation}
In this coordinate frame the ansatz metric $g_{\alpha \beta}$ has
many non--diagonal terms which complicates the study of this
metric in the coordinate frame. In contrast in the
anholonomic frame the metric has only diagonal terms (see
Eq.(\ref{ansatz4}) which makes the system much easier to study.
From the above coordinate frame form of the ansatz metric one
can see that in the locally isotropic limit --
$w_{2,3}\rightarrow 0$ , $n_2\rightarrow \omega \left( r\right)$
and $n_3\rightarrow n\cos \theta$ -- the DS--solution is recovered.
For small anisotropies the solutions presented above are similar to
the five classes of wormhole or flux tube solutions enumerated at
the end of section II. Stronger anisotropic, vacuum, gravitational
polarizations could result in nonlinear renormalization of the effective
``electromagnetic'' constants, and could change substantially the wormhole
-- flux tube configurations.

\section{Example :\ Anisotropic warped wormhole like
configuration }

Up to this point we have implicitly been working in the
standard Kaluza-Klein scenario where the extra dimension,
$\chi$ is assumed to be ``curled up'' or compactified.
However, within the present set up it is possible to find
wormhole like solutions which treat the 5$^{th}$ coordinate
as in the Randall-Sundrum (RS) scheme \cite{rs} where an
exponential warp factor multiples the four
other coordinates ({\it i.e.} the RS metric has a form like
$ds^2 = e^{-2k|\chi |}\eta _{\mu \nu} dx^{\mu} dx^{\nu}
+d \chi ^2$ which exponentially suppresses motion off of
the 4D space). To this end we multiple the d--metric ansatz
(\ref{ansatz3}) by the conformal factor
$\Omega _0^2(\zeta,\theta) \exp[-2 k_\chi |\chi |]$
where $k_\chi $ is an arbitrary constant
\begin{equation}
\label{ansatz6}
\delta s^2 = \Omega _0^2(\zeta,\theta) 
\exp(-2k_\chi  |\chi|) \left[ dt^2- b(\zeta) d\zeta ^2-
 c(\zeta)d\theta ^2 -h_4\left(\zeta,\theta, \chi \right) 
(\breve{\delta} \chi)^2 -  h_5\left( \zeta,\theta, \chi\right) 
(\delta \varphi)^2 \right] \nonumber
\end{equation}
where for the local coordinates are chosen
$x^1= t ,  x^2=\zeta , x^3=\theta , y^1 = v= \chi ,
y^2= p= \varphi , $ with $\breve{\delta} \chi$ being an
N--elongation of the $\chi$--variable
$$ \breve{\delta} \chi = \delta \chi 
+ {\breve w}_{i}(x^{k}, \chi)dx^i
+ {\breve w}_\varphi(x^{k}, \chi) \delta \varphi,
$$
and $\delta \chi$ and $\delta \varphi$ are still
N--elongated as in (\ref{ndiff2}).

The 5D vacuum Einstein equations for this ansatz reduce to the
system of equations (\ref{einsteq3a})--(\ref{einsteq3d}) 
and some additional
equations for the off--diagonal components of the Ricci
d--tensor,
\begin{equation} \nonumber 
P_{4{\breve i}}= - \psi _\chi \delta _{\breve i} \ln \sqrt{|h_4|},
\end{equation}
where $\psi _\chi = \partial _\chi \ln |\Omega|$ for 
$\Omega = \Omega _0 \exp( - 2k_\chi |\chi|)$ and
$\delta _{\breve i} = \partial _{\breve i} - {\breve w}_{\breve i}$. The
index ${\breve i}$ runs over $i$ and $\varphi$, {\it i.e.}
${\breve w}_{\breve i} = ({\breve w}_{i}, {\breve w}_{\varphi})$.
A solution to these equations is
\begin{eqnarray}
\Omega _0 &=& c^{1/4}(\zeta ) |\sin \theta |, \qquad 
 h_4 = c(\zeta) \sin ^2 \theta \exp( - 4k_\chi |\chi|),\qquad   
 h_5=\exp [2\psi (\zeta )]\widehat{r}_0^2(v)\  ; 
\nonumber \\
{\breve w}_1&=&0,\qquad {\breve w}_2 =
 [\ln|\ln |c^{1/4}(\zeta )|| ]^{\prime},
\qquad {\breve w}_3 = [\ln |\ln |\sin{\theta}|| ]^{*},
 {\breve w}_{\varphi} =0. \label{data6}
\end{eqnarray}
The specific form for the functions $b(\zeta), c(\zeta),$ and
$h_5$ is the same as in Eqs. (\ref{aux10}), (\ref{h5}) and (\ref{linr}).
This d--metric (\ref{ansatz6}) with ansatz functions of the
form (\ref{data6}) has the interesting property that it
admits a warped conformal factor $\exp(-2k_\chi  |\chi|)$
which is induced by a conformal transformation of the anisotropic
wormhole like solution (\ref{ansatz3}).

One of the chief consequences of the RS scheme is an
{\it isotropic} deviation of Newtonian gravitational
potential. In the original RS scenario this takes the form
$$
V(r) = G_N \frac{m_1 m_2}{r} \left( 1+ \frac{1}{r^2 k^2}
\right)
$$
for two interacting masses $m_1 , m_2$. The deviation
is isotropic since it only depends on $r$. Using a set up
similar to our wormhole example of Eqs. (\ref{ansatz6})-
(\ref{data6}), but applying it to the two brane scenario
of Ref. \cite{rs} shows that anholonomic coordinates can
give rise to ``anisotropic'' deviations of the Newtonian
potential like \cite{vbrane}
$$
V(r)=G_N{\frac{m_1m_2}r} \left( 1+\frac{e^{-2 k_y |y| }}
{r^2k^2}\right)
$$
where $y$ is a space-like coordinate and $k_y$ is a constant
to be measured or constrained experimentally. The details of such a
construction can be found in Ref. \cite{vbrane}

\section{Conclusions}

Solutions of wormhole--flux tube type are of fundamental importance in the
construction of non--perturbative configurations in modern string theory,
extra dimensional gravity and quantum--chromodynamics. Any
solutions which aid in the physical understanding of such models are clearly
beneficial. The solutions presented here have a number of
features which achieve this aim. In addition this paper illustrates the
usefulness of the method of moving anholonomic frames in finding new metrics
and new types of locally anisotropic field interactions.

The class of solutions presented here demonstrates that wormholes and flux
tubes are not only spherically symmetric, but can also have an anisotropic
structure. Moreover these solutions demonstrate that the
higher dimensional gravitational vacuum can be
``polarized''. This induces observable anisotropic effects
in 4D spacetime: either an angle dependent interaction from
the anisotropies in $\varphi$, or a running of the charges of
the solution coming from the anisotropies in $\chi$. Such solutions
with generic local anisotropy are consistent, in the locally
isotropic limit, with previously known wormhole metrics
\cite{ds,chodos,dzhsin}. In section V we gave a wormhole like
solution where the 5$^{th}$ coordinate was treated as in
the RS scenario -- motion off the 4D brane into the extra
dimension is suppressed by an exponential warp factor. Our solution
also exhibited an anisotropy through its dependence on the
anholonomic variables. Applying these ideas to the
Newtonian potential would lead to anisotropic deviations
from the classical $1/r$ form \cite{vbrane}

A substantial new result of this paper is that the vacuum Einstein equations
for 5D Kaluza--Klein theory can describe higher dimensional, locally
anisotropic gravitational fields. These fields in turn induce anisotropic
field interactions in 4D gravity and matter field theory. This emphasizes
the importance of the problem of fixing ones reference system in gravity
theories. This is not done by any dynamical field equations, but via
physical considerations arising from the symmetries of the metric and/or the
imposed boundary conditions. The analysis presented here shows that the
anholonomy of higher dimensional frame (vielbein) fields transforms the
`lower' dimensional dynamics of interactions,
basic equations, and their solutions, into
locally anisotropic ones. This in part supports the results of 3D and 4D
calculations \cite{v} that locally anisotropic configurations and
interactions can be handled in general relativity and its extra--dimensional
variants on (pseudo) Riemannian spacetimes, through the application of the
method of anholonomic frames.

\section*{Acknowledgments}

S. V., V. B. and D. D. are grateful to the participants of the Seminar
``String/M--theory and gravity'' from the Academy of Sciences of Moldova for
useful discussions. D.S would like to thank V. Dzhunushaliev for
discussions related to this work.

\end{document}